# From Propulsion to Suction: Unraveling Thrust Reversal in Propellers at Intermediate Reynolds Numbers

Rong Fu [1], Siyu Li [2], Yang Ding [3]*

[1] Division of Mechanics, Beijing Computational Science Research Center; Beijing 100193, China.

[2] School of Physics and Astronomy, Beijing Normal University, Beijing 100875, China.

[3] School of Physical Science and Technology, Beijing University of Posts and Telecommunications; Beijing 102206, China.

*Yang Ding.

**Email:** dingyang@bupt.edu.cn

PNAS strongly encourages authors to supply an ORCID identifier for each author. Do not include ORCIDs in the manuscript file; individual authors must link their ORCID account to their PNAS account at www.pnascentral.org. For proper authentication, authors must provide their ORCID at submission and are not permitted to add ORCIDs on proofs.

**Author Contributions:**

Conceptualization: RF, YD

Methodology: RF, SL

Investigation: RF, SL

Visualization: RF, SL

Funding acquisition: YD

Project administration: YD

Supervision: YD

Writing – original draft: RF, SL, YD

Writing – review & editing: RF, SL, YD

**Competing Interest Statement:** Authors declare that they have no competing interests.

## Abstract

This study investigates propeller hydrodynamics at intermediate Reynolds numbers (*Re*), crucial for small-scale robotic systems but still uncharted. Experiments on a propeller-driven underwater vehicle and numerical simulations reveal thrust reversal—a phenomenon where clockwise propeller rotation leads to backward motion—in the approximate range $1.3 \lesssim Re \lesssim 150$ under specific conditions. Notably, counterclockwise rotation consistently results in backward motion. Simulations reveal that this behavior arises when centrifugal suction, an inward force along the axis caused by radial outward flow from the propeller's rotation, dominates over fluid backward acceleration, the primary thrust mechanism at high *Re*. These findings provide critical insights into the unique dynamics of the intermediate *Re* regime and inform the design of efficient propulsion systems for miniature aquatic robots.

**Significance Statement**

This study reveals a surprising reversal of thrust direction in propellers operating at intermediate Reynolds numbers—a regime highly relevant to small-scale robotic systems and other technologies operating in viscous or low-inertia environments. The discovery challenges conventional understanding of rotation-driven propulsion and reveals that changes in fluid dynamics at intermediate *Re* can fundamentally alter thrust generation. These findings open new directions for designing efficient propulsion strategies across a range of applications, from medical devices to environmental and industrial systems.

## Introduction

Propulsion systems for moving in fluids are essential across various technological domains, with the key component being the mechanism that generates thrust through interaction with the surrounding fluid. The development of screw propellers in the early 19th century, alongside the advent of the steam engine, marked a revolution in propulsion technology by significantly surpassing the efficiency of paddle wheels. Today, propellers are used in a wide array of applications—from submarines navigating ocean depths to helicopters flying on Mars—highlighting their versatility and enduring importance (1, 2).

Successful propeller design hinges on a deep understanding of the fluid mechanics of propulsion. At the macro-scale, extensive research has enabled the development of highly efficient propulsion systems that adapted to both aquatic and aerial environments (3, 4). At the micro-scale, studying the hydrodynamics of nature's own propellers—such as bacterial flagella—has greatly enhanced our understanding of the morphology, behavior, and evolution of microorganisms (5–9). A crucial parameter for understanding these dynamics is the Reynolds number (*Re*), which quantifies the relative importance of inertial to viscous forces. At high *Re* ($\gtrsim 10^3$), thrust is generated from the reaction force exerted by accelerating fluid (3, 4), while in extremely low *Re* environments ($Re << 1$), inertia is negligible, and thrust is generated primarily through anisotropic viscous drag (10). Despite significant advances at both ends of this spectrum, the hydrodynamics of propeller at intermediate *Re* (approximately $10^{-2} < Re < 10^2$) remains uncharted.

The growing interest in a wide range of small-scale and precision applications, particularly miniaturized robots, has amplified the significance of understanding propulsion at intermediate *Re*. In aquatic environments, small-scale devices are developed for pipe inspection, maintenance, and water treatment (11–14). For example, a 2.5mm diameter robot prototype that swims at $Re \approx 32$ was developed to fight pollutants and bacteria with antibiotic and photocatalytic coating (15). Swimmers with size range of hundreds of microns to a few millimeters are especially promising for medical applications, including plaque or clot removal in blood vessels (16, 17), surgical treatment and drug delivery in the eyes (18), urinary tract (19), and gastrointestinal tract (20, 21). In the air, ever smaller insect-inspired flying robots are being developed (22, 23). One day, they may encounter intermediate *Re* regime in which their biological counterparts have navigated for millions of years (24–26). While some miniature robots have achieved navigation using propellers based on high or low *Re* theories, the limited understanding of propeller hydrodynamics at intermediate *Re* significantly hampers the development of efficient and reliable propulsion systems, restricting the full potential of microrobots across diverse applications.

Here we aim to bridge this gap by investigating the performance and hydrodynamics of propellers at intermediate *Re*. Our findings reveal unexpected phenomena that challenge traditional propeller design principles and offer new insights for micro-robot propulsion.

## Results

### Forward Motion Disappearance in a Toy Submarine

We employ a 75mm long remote toy submarine with a rear-mounted propeller (Fig. 1A–C). Throughout this study, we define clockwise (CW) and counterclockwise (CCW) rotation as viewed from behind the submarine. The *Re* is defined as ${D_p}^2n/\nu$, where $D_p$ represents the diameter (5mm) of the submarine's propeller, *n* denotes the rotation speed of the propeller in unit of rotations per second, and $\nu$ signifies the kinematic viscosity of the fluid. In water, where *Re* > 7500, the submarine's movement direction aligns with the propeller's rotation—it moves forward with CW rotation and backward with CCW rotation. To study the submarine's motion at intermediate *Re*, we immersed it in dimethyl silicone oil with varying viscosities (Fig. 1D, see materials and methods for details). By using silicone oils of higher viscosity, we reduced *Re* to the range of 0.4–350, enabling the investigation of hydrodynamic behavior in the intermediate *Re* regime.

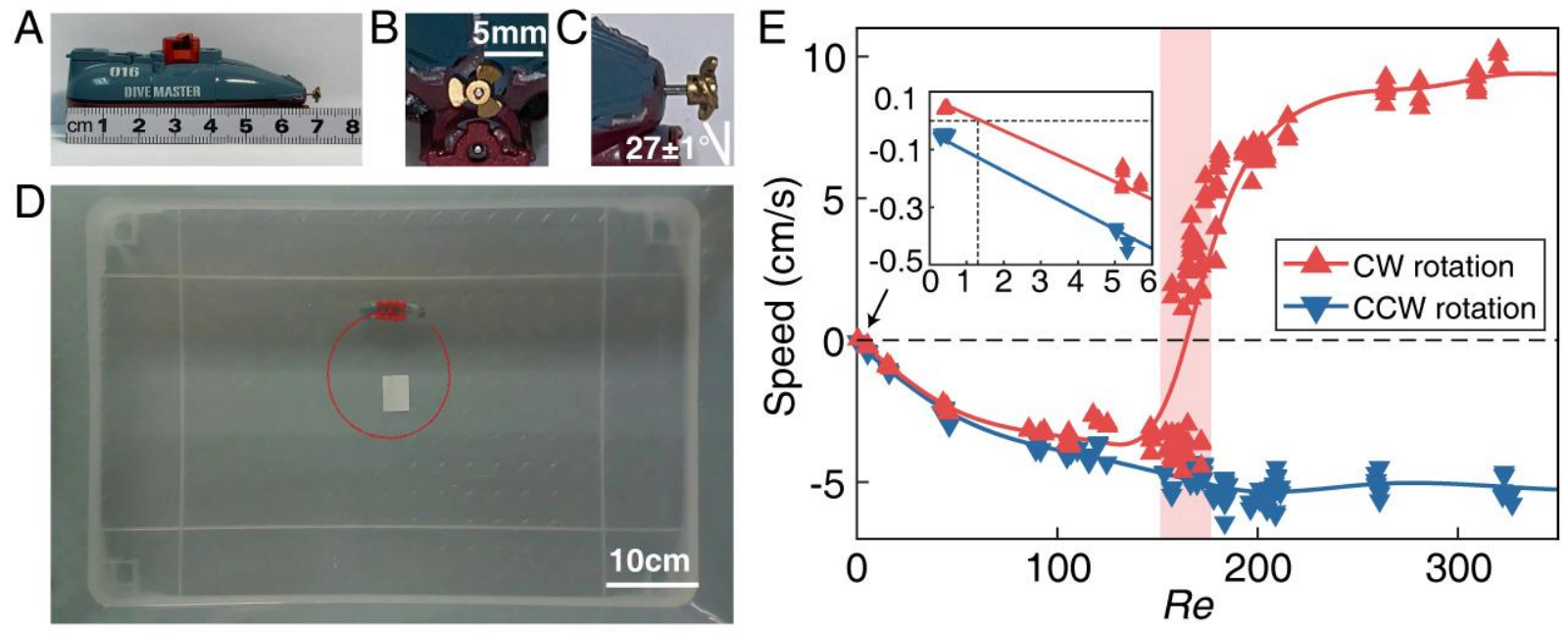

**Fig. 1. Movement of toy submarines in silicone oil.** (**A**) A miniature toy submarine (75 mm length, 20 mm width, 30 mm height) with a right-handed 5 mm propeller at the stern. (**B, C**) Close-up views of the three-blade propeller. (**D**) Representative snapshot of the toy submarine's motion; red line shows trajectory. (**E**) Speed vs Reynolds number for CW (red) and CCW (blue) propeller rotation. Inset: enlarged view of the data in the small-*Re* range. Curves are spline fits. The red shaded region shows the range of *Re* where CW rotation can result in both forward and backward motion. The horizontal dashed line denotes zero speed. The vertical dashed line marks the interpolated *Re* ≈ 1.3 at which motion under CW propeller rotation reverses. Fluid viscosity: $\nu$=18.6-1900.7 mm$^2$/s. The corresponding rotation speeds (*n*) for these data are provided in Fig. S14.

At high Reynolds numbers (*Re* ≳ 175), the behavior of the toy submarine is similar as in water: CW propeller rotation drives forward motion, while CCW rotation results in backward motion, consistent with its behavior in water (Fig. 1E). As *Re* decreases into the range *Re* ≈ 150–175, the direction of motion becomes unstable—sometimes forward, sometimes backward—depending sensitively on initial conditions, as indicated by the red shaded region in Fig. 1E. Below this range, in *Re* ≈ 1.3–150, the submarine exhibits a clear thrust reversal: CW rotation now produces backward motion, contrary to the high *Re* behavior, while CCW rotation continues to result in backward motion. Finally, at *Re* ≲ 1.3, a second reversal occurs—CW rotation once again leads to forward motion, though at low speed, while CCW rotation still drives backward motion (Fig. 1E, inset).

### Thrust Reversal in a Fixed Propeller System

To further investigate the causes of the unexpected reversals in the movement of the toy submarine in the CW rotation case, we design a simplified propeller system, hold it in place, and study the static thrust (breakaway force) (Fig. 2A). The simplified propeller system consisted of a symmetric propeller and a disk, where the disk represented the "body" of the submarine. The rationale for including the ‘body’ is as follows: Despite the near symmetry of the propeller, the toy submarine exhibited asymmetric motion, i.e. it moves backward regardless of the propeller's rotation direction. As such, the asymmetric positioning of the body of the submarine relative to the propeller likely plays a significant role. In the experiment, we varied the propeller's rotational speed and measured the net axial force ($F_{net}$) on the propeller and disk at steady state (see materials and methods for details). For consistency with the toy submarine’s movement, we defined the direction from the propeller to the disk as the positive force direction. All forces were non-dimensionalized using a characteristic inertial force, $F_0 = \rho {D_p}^4 n^2$, where $\rho$ represents the density of the silicone oil.

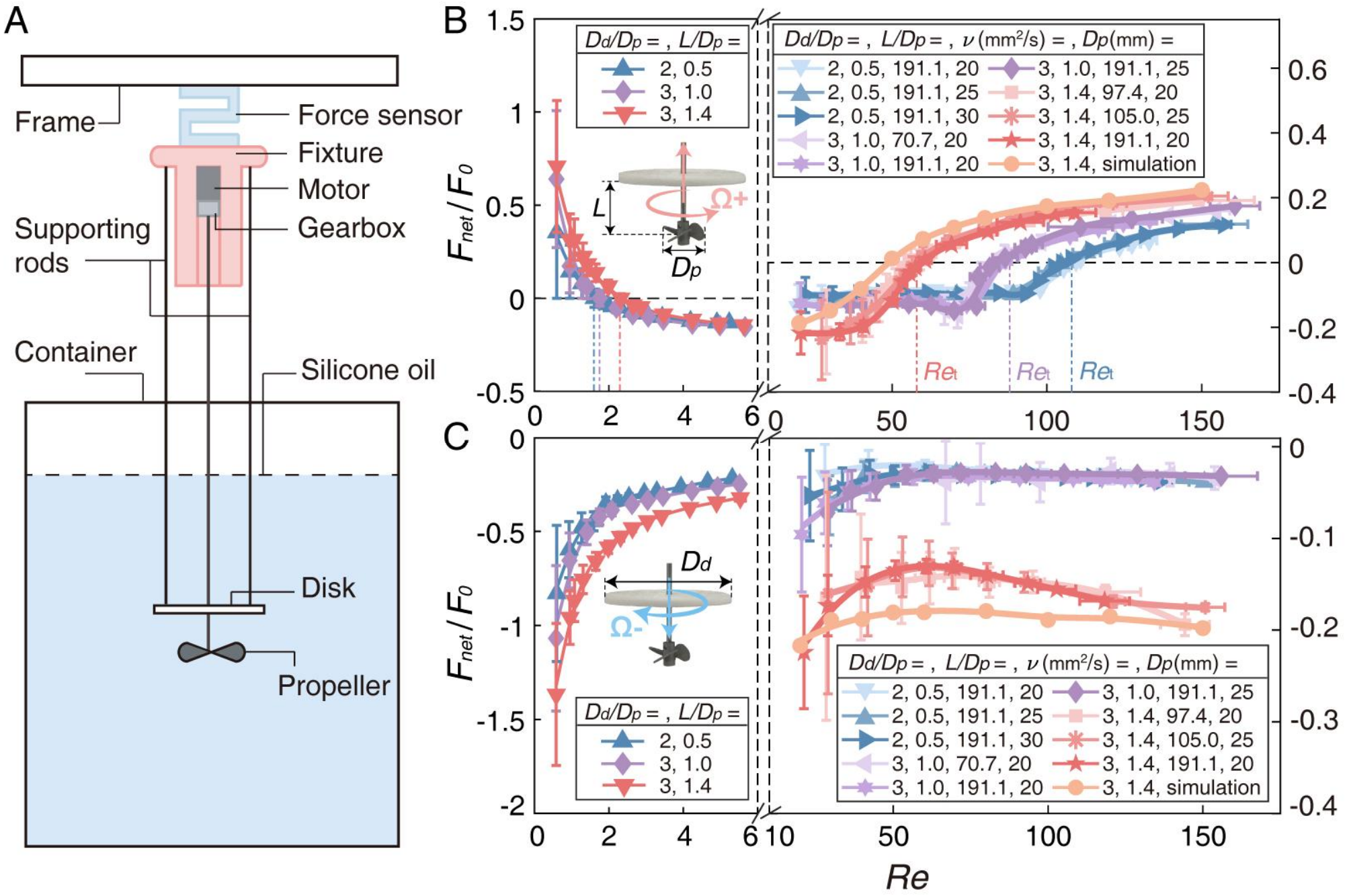

**Fig. 2. Net force on the simplified propeller system across varying geometric and fluid parameters.** (**A**) Experimental setup: Forces on the disk and propeller are transmitted vertically to a force sensor via supporting rods. (**B, C**) Net force as a function of Reynolds number during CW (B) and CCW (C) rotation. Dynamic viscosity ($\nu$, in mm²/s) of the silicone oil, propeller diameter ($D_p$, in mm), diameter of the disk ($D_d$, in mm), and distance between the propeller and the disk ($L$, in mm) were varied in the experiments. The low-*Re* range (left panels of B and C) corresponds to $\nu$ = 2804.1 mm²/s and $D_p$ = 25mm. For comparison, simulation with the parameter $D_d / D_p$ = 3 and $L / D_p$ = 1.4 are represented as orange dots (see Materials and Methods for details). Solid lines represent spline fits to the data, serving as visual guides. Dashed horizontal lines indicate zero net force, and dash-dotted vertical lines denote the interpolated Reynolds numbers where the net force crosses zero. Insets show the simplified propeller system in experiments, key geometric parameters, and the direction of propeller’s rotation.

The measured forces reproduce the key features observed in the moving system (Fig. 2 B and C). For CW rotation, $F_{net}$ changes sign twice as *Re* decreases—first from positive to negative at an intermediate Re (hereafter denoted $Re_t$), then back to positive at very low *Re*. For CCW rotation, the force remains negative throughout. This thrust reversal phenomenon in the CW case— alongside the unidirectional force in the CCW case—is robust and reproducible across propellers with varying blade numbers and pitch angles (Fig. S1, S2).

Our focus in this study is on the first reversal at intermediate *Re***,** where the flow is more intricate and the cause of the transition is less apparent. At *Re* = 0, the Stokes equations are strictly time-reversible, meaning that CW and CCW rotation must produce exactly opposite motion or thrust. Consequently, if CW rotation yields negative thrust at some finite *Re*, there must be a second reversal to positive thrust as *Re* approaches zero. Understanding this reversal also clarifies the low-*Re* case, which can be understood as the disappearance of the same underlying mechanism.

To identify the parameters controlling the system's behavior, we systematically varied propeller diameter ($D_p$), disk diameter ($D_d$), distance between propeller and disk ($L$), rotation speed ($n$), and fluid viscosity ($\nu$). When $D_d$ and $L$ are normalized by $D_p$, the $F_{net}$ - $Re$ curves collapse into master curves for specific combinations of $D_d / D_p$ and $L / D_p$ values (Fig. 2, B and C), indicating these as the primary non-dimensional control parameters.

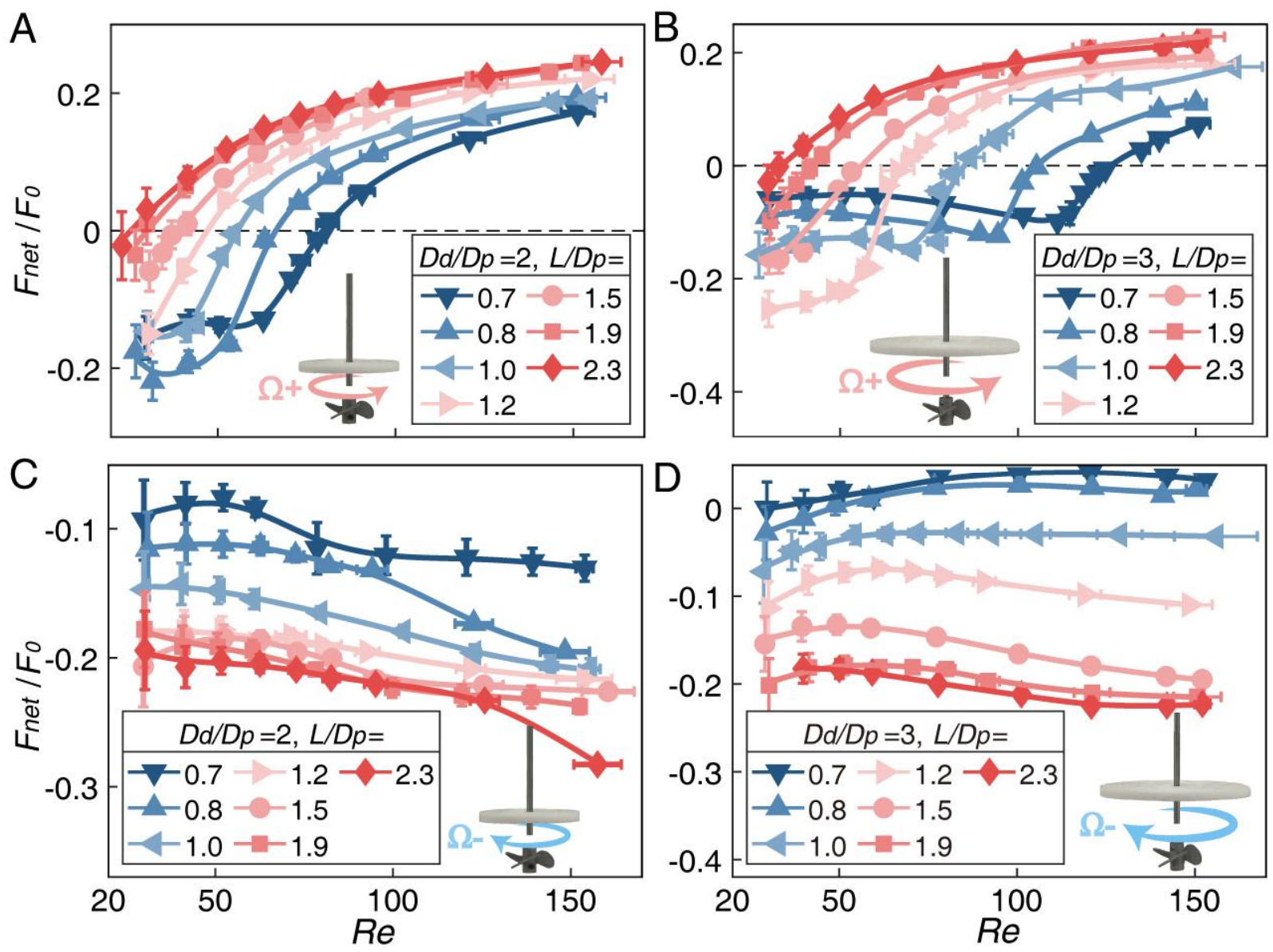

**Fig. 3. Influence of disk diameter ($D_d$) and distances ($L$) on the net force on the simplified propeller system with fixed propeller diameter ($D_p$).** $D_p$ = 25 mm and $\nu$ = 191.1 mm$^2$/s in all cases. Left column: $D_d$ = 50 mm, propeller rotated in the CW (**A**) and CCW (**C**) direction. Right column: $D_d$ = 75 mm, propeller rotated in the CW (**B**) and CCW (**D**) direction. Insets show the direction of rotation of the experiment system.

By varying two of the primary parameters—$D_d / D_p$ and $L / D_p$, we noted significant changes in the net force during CW rotation (Fig. 3, A and B). Specifically, increasing the distance reduced the $Re_t$. Additionally, smaller disk sizes resulted in lower $Re_t$ for the same distance (see the same colors in Fig. 3, A and B). These observations suggest that the disk's obstruction of the flow along the axis contributes to the net force reversal in CW rotation. For CCW rotation, smaller distances and larger disk sizes led to smaller force magnitudes (Fig. 3, C and D). The effect of *Re* on the variation of the net force was relatively less pronounced compared to the CW rotation cases.

## Transitions from backward fluid acceleration to centrifugal suction

To elucidate the underlying mechanism of the observations on the net forces, particularly the reversal of thrust in the CW rotation case, we employed 3D computational fluid dynamics simulations that match the experiments on the simplified system (Fig. 4A). These simulations allowed us to examine the flow, pressure, and individual forces on the propeller and the disk (see materials and methods) (27).

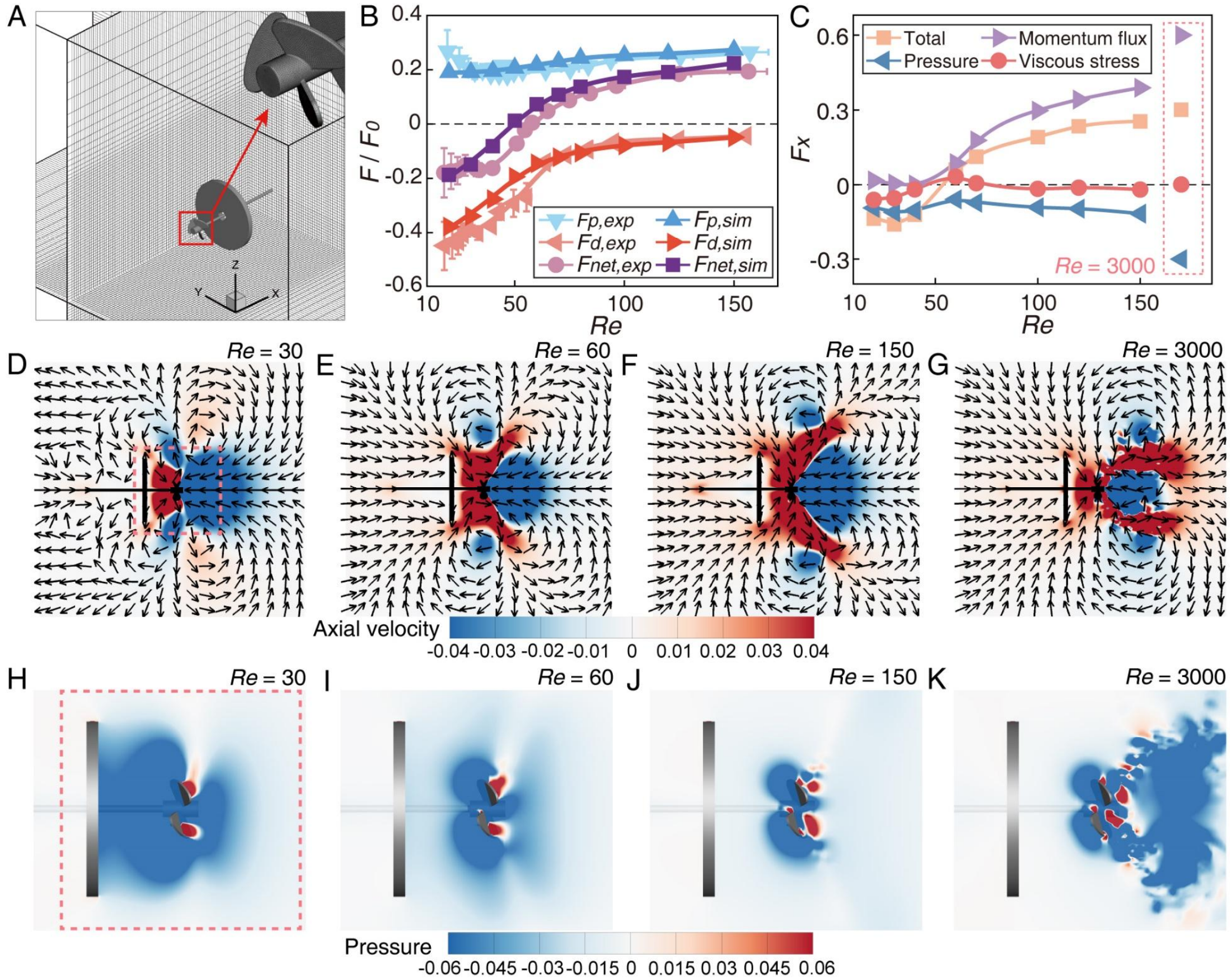

**Fig. 4. Numerical Simulation. (A)** Schematic of the computational model. Non-dimensionalized parameters: propeller diameter $D_p$ = 1.0, disk diameter $D_d$ = 3.0, distance between them $L$ = 1.4, and rotation speed $n$ = 1 (see Materials and Methods for details). The computational domain measures 13.6 × 22.7 × 22.7 and the grid consists of 969 × 935 × 935 nodes. For clarity, only every tenth grid nodes are shown in each of the three dimensions. The red square region is magnified and shown as an inset for clarity. **(B to K)** Results for the CW rotation case: (B) Forces on the disk $F_d$, the propeller $F_p$, and combined $F_{net}$ as a function of *Re* from simulation and experiment. (C) Contributions of momentum flux, pressure, and viscous stress to the axial force, obtained from control volume analysis on the region outlined by pink dashed lines in (D) and (H); their sum is also shown for comparison. (D to G) Flow fields at the y = 0 plane at different *Re*. Black arrows indicate the velocity direction and colors (color bar, inset) representing the velocity along the axial direction. (H to K) Pressure fields in the z = 0.2 plane at different *Re*.

The flow fields in the CW rotation cases clearly show a qualitative transition as Reynolds number decreases (Fig. 4, D-G). At high *Re*, the fluid accelerated by the propeller forms a jet behind the propeller (Fig. 4G). When *Re* decreases, the overall jet flow transitions to a jet with an angle to the axis and an inward flow forms behind. As *Re* further decreases, the angle increases and the outward velocity from the propeller becomes predominant along radial direction and the inward flow from behind becomes strong. These changes suggest a qualitative shift in the way the propeller interacts with the surrounding fluid, coinciding with the range in which thrust reversal is observed.

Along with the transition in the flow field, a shift in the pressure field is noted (Fig. 4, H-K). For most *Re* explored in the simulations, the local pressure on the two sides of a blade aligns with the direction of the propeller rotation: CW rotation generates positive pressure on the backside and negative pressure on the front side, and vice versa for CCW rotation. These pressure distributions result in forces on the propeller in the same direction as the rotation (Fig. 4B & Fig. S3), except that positive forces, relatively smaller, were observed in CCW rotation when $Re \lesssim 30$. At a slightly larger scale, the propeller creates a negative pressure region around it (Fig. 4, H-K). As *Re* decreases, this region extends toward the disk. This expansion coincides with a rapid increase in the magnitude of negative forces on the disk. In the CW rotation case, this increase causes the magnitude of the negative force on the disk to surpass the positive force on the propeller, leading to a transition of the net force from positive to negative. Similarly, an expansion of the negative pressure region and an increase in the negative force on the disk were observed for CCW rotation (Fig. S3).

To further quantitatively connect the observed changes in flow and pressure to net thrust generation, we applied a control volume analysis enclosing the propeller–disk system and decomposed the net axial force into contributions from momentum flux, pressure, and viscous stress (see Materials and Methods). At high *Re*, the dominant contribution comes from momentum flux exiting the control volume, consistent with jet propulsion driven by backward fluid acceleration. As *Re* decreases, the momentum flux declines and becomes negligible, while the contribution from negative pressure—i.e., suction—becomes dominant (Fig. 4C).

Based on the observed changes in flow structure, pressure distribution, and control volume analysis, we hypothesize that the competition between axial suction caused by centrifugal force—referred to as "centrifugal suction"—and backward fluid acceleration is responsible for the thrust reversal. This explains why the net force becomes negative regardless of the direction of rotation, as centrifugal force is independent of the rotation direction. It also explains the increase in $Re_t$ observed in experiments when the disk is closer or larger, as the suction effect becomes stronger in these scenarios.

**Reconstructing the transition from centrifugal suction and backward fluid acceleration**

To further test the hypothesis, we used numerical simulation to isolate these two mechanisms and obtained the respective net forces, denoted as $F_{net,s}$ for centrifugal suction and $F_{net,a}$ for backward fluid acceleration (Fig. 5). Subsequently, we reconstructed the net force ($F_{net,t}$) as the sum from the two mechanisms, i.e., $F_{net,t} = F_{net,s} + F_{net,a}$, and compared them to the net forces directly measured in previous stationary experiments and simulations.

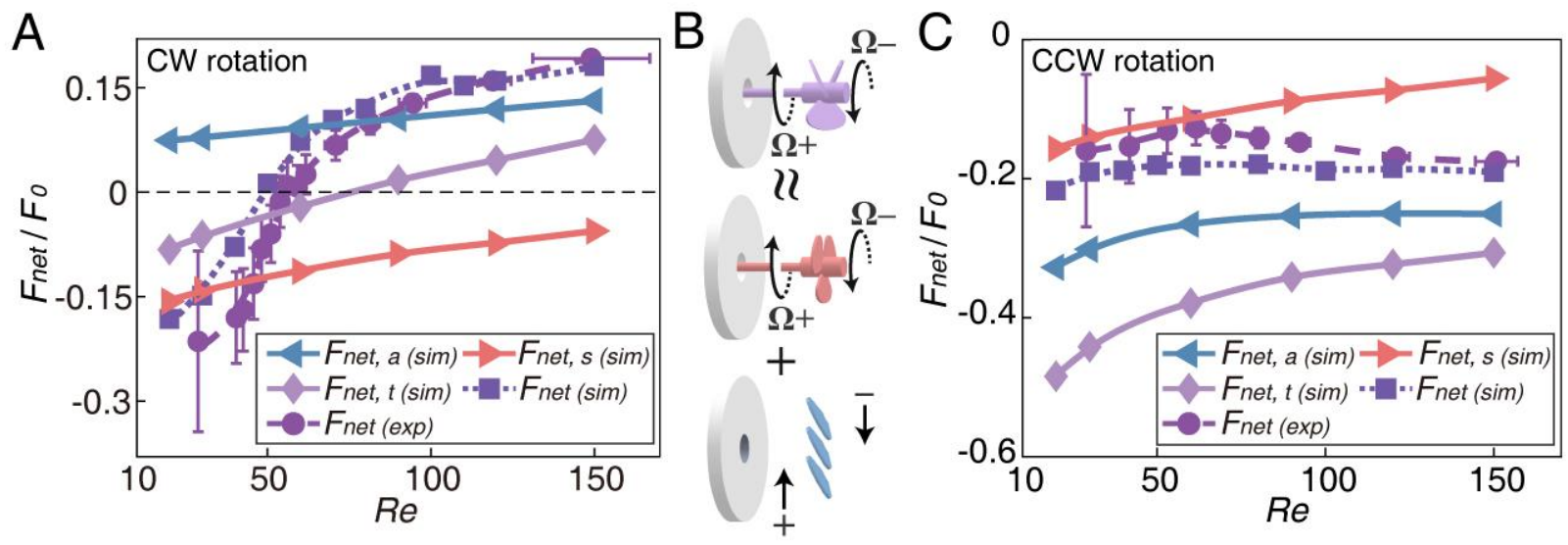

**Fig. 5. Reconstruction of the net force on the simplified propeller system. (A)** Net forces generated by two mechanisms during CW rotation: centrifugal suction ($F_{net,s}$, red) and backward fluid acceleration ($F_{net,a}$, blue), along with their sum ($F_{net,t}$, purple). These are compared with net forces directly measured from simulations and experiments ($F_{net}$). **(B)** Schematic illustrating the decomposition of the net force on a rotating propeller (purple) into the two mechanisms: centrifugal suction (red) from a rotating propeller with zero pitch angle and backward fluid acceleration (blue) from three angled blades translating linearly. Symbols and signs on the left (+) and right (-) indicate the directions of motion for CW and CCW rotation, respectively. **(C)** Same comparisons as in (A) for the CCW rotation case. The color scheme is consistent across all subfigures. Simulation parameters: $D_p$ = 1.0, $D_d$ = 3.0, $L$ = 1.4, and $n$ = 1.

To model the contribution from pure centrifugal suction, the pitch angle of the propeller was set to zero (Fig. 5B, red color). The effect of pure backward fluid acceleration was estimated by replacing the rotational motion of the blades with translational motion (Fig. 5B, blue color). In this configuration, three angled blades move linearly, parallel to the disk, in either the forward or backward chordwise direction. Their distance and speed match those of blades on the rotating propeller: the area-averaged center of each blade is placed at $D_p$ / 3 from the propeller's center, giving a circumferential spacing of $2\pi D_p / 9$, and the blades' average linear speed is $2n\pi D_p / 3$.

$F_{net,t}$ qualitatively agrees with the net forces from the direct measurements (Purple lines in Fig. 5, A and C). For both rotation directions, the centrifugal suction consistently generates a negative $F_{net,s}$, whose magnitude increases considerably as $Re$ decreases. The sign of $F_{net,a}$ depends on the direction of linear motion, while its magnitude exhibits relatively small changes with $Re$. Consequently, $F_{net,t}$ transitions from positive to negative in the CW rotation case and remains negative in the CCW rotation case.

While this decomposition does not quantitatively reproduce the measured forces, its aim is to qualitatively validate the hypothesized mechanism by showing that centrifugal suction alone can generate a force large enough to reverse the thrust. Deviations between $F_{net,t}$ and the measured values are expected, given the simplified geometry and the absence of interaction between the two mechanisms in the reconstruction. Capturing the full quantitative behavior would require a more detailed, coupled model.

## Discussion

In summary, we discovered an unexpected phenomenon of thrust reversal in a propeller-driven vehicle at intermediate Reynolds numbers. This behavior was corroborated through stationary experiments and numerical simulations, indicating a fundamental shift in propeller hydrodynamics that challenges the established understanding of propulsion mechanisms. Specifically, we observed that centrifugal suction, often neglected at high $Re$, becomes significant enough at intermediate $Re$ to compete with the backward fluid acceleration generated by angled propeller blades. This competition ultimately leads to a considerable reduction in forward thrust, and in some cases, complete reversal.

Our understanding of these competing mechanisms offers several immediate strategies to regain net thrust. One obvious solution is to reduce the suction force by increasing the distance between the propeller and the body. However, the efficiency is likely low as considerable energy is wasted in producing radial flows (Fig. 4C and Fig. S15B). Alternatively, flat-blade or no-blade designs could enhance centrifugal suction by eliminating backward fluid acceleration, thereby preventing force cancellation and achieving backward locomotion. However, these designs may have limited applications due to the unidirectional motion.

To go beyond merely recovering thrust and replicate the success of high-*Re* propellers or low-*Re* flagella, it is essential to develop propellers and similar devices that efficiently convert rotational motion into translational motion. Such development demands guidance from new theoretical frameworks akin to lifting surface/line theories at high *Re* (4) and resistive force theory at low *Re* (10). The findings from our study lay the groundwork for these new theories by identifying key mechanisms that must be considered and providing valuable test cases to validate these models.

In a broader context, our findings highlight the intricate challenges and distinctive hydrodynamics of propulsion in the intermediate Reynolds number regime. Accumulating evidence—from studies of undulatory swimming (28–30), rowing (31–33), flapping (25, 34–39), jet propulsion (40, 41), metachronal paddling (42–45), and other modes (46–56)—has progressively sharpened our understanding of this regime. Among these studies, some have revealed unusual phenomena within this regime—for example, tiny flapping insects adopt hairy wing morphologies (25), and insect larvae employ gaits distinct from those of typical undulatory swimmers (50, 51). Other studies using idealized models have found reversals in swimming speeds as Re varies (51–54).

Building on this body of knowledge, our study not only uncovers a double-reversal phenomenon with two distinct Re transitions, but—more fundamentally—identifies a previously unrecognized mechanism: centrifugal suction, arising purely from flow reorganization induced by changes in Re. Moreover, unlike previous idealized models (e.g., (51, 53)), the propeller—a centuries-old, rotation-driven technology—combines proven robustness with exceptional adaptability, making it uniquely suited to meet the demands of the rapidly evolving field of robotics.

As we deepen our understanding of locomotion at intermediate *Re*, there remains much to discover both in how organisms have evolved to navigate these complexities and in how these insights can inform the development of innovative, efficient propulsion systems for a variety of environments.

## Materials and Methods

### Speed Measurement of Toy Submarines

Three identical toy submarines (75 mm length, 20 mm width, 30 mm height) with rims around the propeller removed were used (Fig. 1A–C). The submarines were fully charged before each experiment to ensure consistent motor performance. Experiments were conducted in dimethyl silicone oil with viscosities ranging from 18.6 to 1900.7mm$^2$/s, prepared by mixing oils of different viscosities. A rectangular container (69 cm × 47 cm × 17 cm) filled with 5 cm depth of silicone oil was used to minimize boundary effects (Fig. S13). The submarine's motion was controlled remotely, with the propeller rotating CW or CCW. Each Reynolds number group ($5 \leq Re \leq 350$) included six velocity measurements: three for CW and three for CCW rotation. Submarine trajectories were recorded using a high-resolution camera (1920 × 1080 pixels, 30 fps) and analyzed using computer vision techniques (Fig. 1D). Velocity data were smoothed using a Savitzky-Golay filter to reduce noise (Fig. S4).

### Rotational speed and axial force measurements

To determine propeller rotational speed, the submarine was fixed in a transparent container filled with silicone oil with $\nu$ = 70.7, 97.4, 105.0, 191.1, 2804.1mm$^2$/s, and high-speed imaging (1200 fps) tracked a pre-marked propeller blade (Fig. S11, Table S1 for propeller parameters). Rotation angles were extracted using OpenCV-based tracking, and water tunnel experiments confirmed consistent speeds between fixed and free-motion conditions (Fig. S5, Supplementary Text). Axial forces on the propeller system were measured using an ARIZON-8101 force sensor (±0.02% accuracy) with a high-precision controller, recording forces once stabilized (Fig. S7). Specialized fixtures connected components to the sensor for individual force measurements (Fig. S6 and Fig. S10). Systematic tests confirmed that boundary effects were negligible, with the container filled to a fluid depth of 34 cm and a diameter of 63 cm being sufficiently large (Fig. S8, S9).

**3D Computational Fluid Dynamics Simulation**

Three-dimensional CFD simulations were performed using an in-house immersed boundary method coupled with a finite-difference scheme (Fig. 4A). The computational domain measured 13.6 × 22.7 × 22.7 and was discretized into 969 × 935 × 935 grid nodes. The propeller and disk were modeled in SOLIDWORKS (SolidWorks Corporation) and meshed using triangular surface elements in HyperMesh (Altair). Simulations were conducted in a non-dimensional framework, with $D_p$ = 1.0 as the reference length and the rotation period as the time scale (i.e., $n$ = 1). *Re* was varied systematically by adjusting the kinematic viscosity $\nu$. We applied no-slip boundary conditions on all domain boundaries and on the surfaces of the propeller system. Forces were averaged over the final full revolution to ensure steady-state behavior (Fig. S12). A fixed time step of $5 \times 10^{-4}$ was used. Grid independence and time-step sensitivity tests were conducted to verify the numerical accuracy of the results.

In the control volume analysis, we selected a cylindrical control volume Ω centered on the propeller axis, with diameter and length both equal to $4D_p$, enclosing both the propeller and the supporting disk (pink dashed box in Fig. 4D, H). The net axial force exerted by the fluid is computed using the momentum conservation equation:

$$F_x = \frac{d}{dt}\int_{\Omega} \rho u_x \, dV + \int_{\partial\Omega} \rho u_x (\boldsymbol{u}\cdot\boldsymbol{n}) \, dA + \int_{\partial\Omega} p n_x \, dA - \int_{\partial\Omega} \rho\nu \left(\frac{\partial u_x}{\partial x_j} + \frac{\partial u_j}{\partial x}\right) n_j \, dA,$$

where $\boldsymbol{u}$ is the fluid velocity, $\boldsymbol{n}$ is the unit outward normal vector on the control surface, and $p$ is the pressure. As the system reaches steady state, the time-dependent term becomes negligible, and the net axial force is dominated by the momentum flux, pressure forces, and viscous stresses across the control surface. The integrals are evaluated using the CFD simulation results, with surface quantities interpolated from the nearest computational grid points.

**Acknowledgments**

We thank Prof. Li-Shi Luo for helpful discussions and suggestions, and Prof. Zhiguo Wang and Prof. Jun Zhang for carefully reading the manuscript and providing constructive feedback. We also thank Ning Shen for valuable assistance with manuscript preparation. Various publicly available AI-assisted language editing tools were used during the preparation of the manuscript to improve clarity and readability.